\documentclass[useAMS,usenatbib]{mn2e}

\usepackage{psfig}
\usepackage{amsmath}
\usepackage{amssymb}
\usepackage{upgreek}
\usepackage{longtable}
\usepackage[mathscr]{eucal}
\usepackage{graphicx}
\usepackage{float}
\usepackage{subfig}
\usepackage[export]{adjustbox}
\usepackage[usenames, dvipsnames]{color}
\usepackage{titlesec}
\titlespacing*{\section}{0pt}{1.1\baselineskip}{\baselineskip}
\titlespacing*{\subsection}{0pt}{1.1\baselineskip}{\baselineskip}

\title
[The low mass population of the Vela OB2 association from Gaia]
{The low mass population of the Vela OB2 association from Gaia}
\author
[Armstrong et al.]
{Joseph J. Armstrong$^{1}$, Nicholas J. Wright$^{1}$ and R. D. Jeffries$^{1}$\\
$^{1}$Astrophysics Group, Keele University, Keele, ST5 5BG, UK
}

\begin{document}
\maketitle

\begin{abstract}
The first Gaia Data Release presents an opportunity to characterise the low-mass population of OB associations, providing larger statistical samples and better constraints on the formation and evolution of clusters and associations. Using previously known low mass members in a small region of Vela OB2 we have designed selection criteria that combine Gaia and 2MASS photometry, independently of any astrometric information, to identify low-mass pre-main-sequence (PMS) stars over the wider association area. Our method picks out the known clusters of young stars around $\gamma^2$ Velorum and NGC-2547, but also identifies other over-densities that may represent previously unknown clusters. There are clear differences in the spatial distributions of the low-mass and the high-mass OB populations, suggesting either that the structure and dynamics of these populations has evolved separately or that the initial mass function can vary considerably on small scales within a single association.
\end{abstract}


\section{Introduction} OB associations are gravitationally unbound groups of OB stars sharing a common motion through space \citep{blaauw64}. They make attractive subjects for studies of star formation and evolution due to their proximity to regions of ongoing star formation. Data from the Hipparcos satellite \citep{perryman97} has allowed the study of the kinematics of bright stars in these associations \citep{dezeeuw99}, but these are only the most massive members and a small fraction of the total stellar population, assuming it follows a classic Initial Mass Function (IMF). Much is obscure about the population of low-mass stars in associations. To what extent do their positions and velocities correlate with those of the OB stars, how do they evolve, and which processes are responsible for this evolution?  \\
\indent Several hypotheses have been proposed to explain the origins of OB associations. The foremost being that associations are the remnants of originally dense, but since dispersed star clusters. They are produced when young O- and B- type stars sweep away their parental molecular cloud, and from this loss of binding mass, the newly formed stars become gravitationally unbound \citep{hills80,lada84,lada03}. However, recent kinematic studies have not found the radial expansion patterns predicted by such models \citep{wright16,wright18}. Another hypothesis is that associations are made up of a continuous distribution of stars at both low and high density, such that high-density regions collapse to form bound clusters while low-density regions disperse as unbound associations \citep{kruij12}. To account for associations often being composed of smaller subgroups of differing ages, \citet{elmegreen77} proposed a sequential formation model where an ionization shock from newly formed OB stars is driven into a molecular cloud, compressing material and triggering a new generation of star formation. This new group of stars can begin the process over again in a chain reaction, until the cloud is dispersed and the association is left. \\
\begin{figure}
\includegraphics[width=\columnwidth]{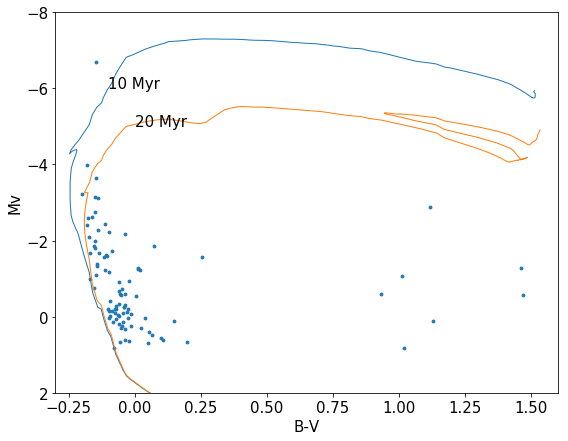}
\setlength{\belowcaptionskip}{-10pt}
\setlength{\textfloatsep}{0pt}
\caption{B-V vs Mv colour-magnitude diagram of the 93 high-mass association members from \citet{dezeeuw99} using B-V and V from the Extended Hipparcos Compilation of \citet{anderson12}, parallaxes for 86 stars from Gaia DR2 \citep{luri18} and dereddened by $A_V = 0.131$ mag \citep{jeffries14}. For the 7 high-mass stars without DR2 parallaxes, the median parallax value of the sample was used. Also plotted are 10 and 20 Myr isochrones from \citet{ekstrom12} for stars with $Z=0.014$, using bolometric corrections (BCs) for B type stars from \citet{pecaut13}.}
\label{KSbes}
\end{figure}
\indent An up-to-date picture of a nearby association, making use of the newest data to identify as much of its population as possible, may lead to a clearer understanding of association evolution and the determination of a likely formation scenario. Vela OB2 is an association at a mean distance of $410\pm12$ pc, containing 93 OB stars based on Hipparcos positions and proper motions spread over an area of $\sim100$ deg$^2$ \citep{dezeeuw99}. The brightest object in the association is $\gamma^2$ Velorum, a spectroscopic binary composed of a massive O star and a Wolf-Rayet star\citep[$28.5 \pm 1.1 M_{\odot}, 9 \pm 0.6 M_{\odot}$,][]{north07}, projected against the centre of the association. The $\sim1$ square degree around $\gamma^2$ Velorum is known to contan a rich population of low-mass pre-main sequence (PMS) stars \citep{pozzo00,jeffries09,jeffries14}. \\
\indent  In this letter we use Gaia Data Release 1 \citep{gaia16a,gaia16b} and 2 Micron All-Sky Survey \citep[2MASS][]{cutri03} photometry to identify the wider low-mass stellar population of Vela OB2, with the aim of investigating its structure, traced by the more numerous low-mass stars. We use the detailed spectroscopic investigation of stars immediately around $\gamma^2$ Vel \citep{jeffries14,prisinzano16} to define photometric selection criteria capable of identifying low-mass PMS stars with minimal contamination. These criteria are applied to the much wider Vela OB2 region to map the spatial distribution of the low-mass population and compare it to that of the high-mass members. 
\section{Data Analysis}
\begin{figure*}
\begin{center}
    \subfloat{{\includegraphics[width=235pt]{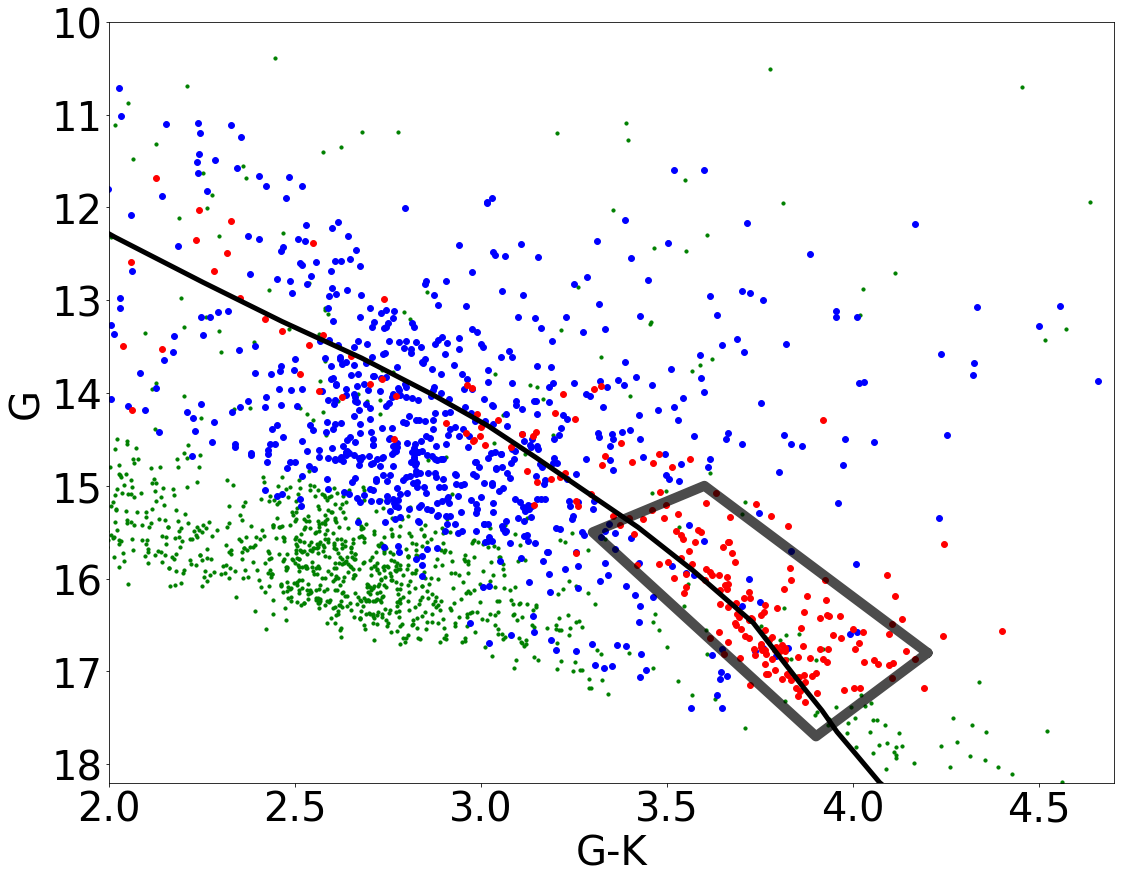} }}%
    \qquad
    \subfloat{{\includegraphics[width=235pt]{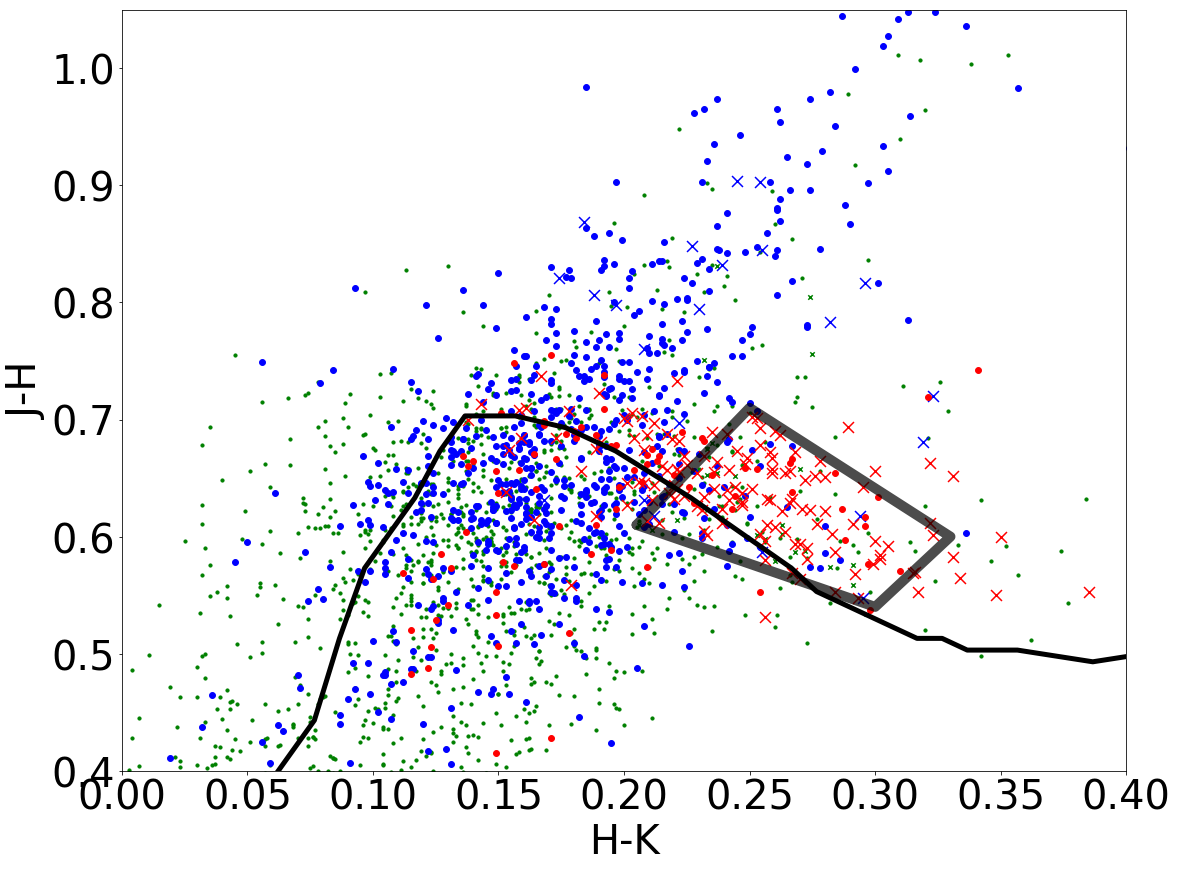} }}%
    \setlength{\belowcaptionskip}{-10pt}
    \setlength{\textfloatsep}{0pt}
    \caption{$Left:$ G-K vs G colour magnitude diagram of objects with both Gaia and 2MASS photometry (green) within a $0.9$ degree$^2$ area around $\gamma^2$ Velorum, spectroscopically identified young stars (red) and field stars (blue) \citep{jeffries14,prisinzano16}. Also shown is a 10 Myr PMS isochrone from \citet{baraffe15} at a distance modulus of 7.76 and reddened by $A_V = 0.131$ mag, as well as our selection box for PMS stars. $Right:$ H-K vs J-H colour-colour diagram for the same objects. Objects located inside the G-K vs G box are marked with crosses. A 10 Myr PMS isochrone from \citet{baraffe15} and a second selection box are shown.}%
    \label{SelectionBoxes}%
\end{center}
\end{figure*}
\subsection{Estimating the size of the low-mass population of Vela OB2} In order to estimate the total stellar mass and expected low-mass population of Vela OB2, we created model populations using the \citet{maschberger13} IMF (with lower mass limit $0.01 M_{\odot}$, upper mass limit $150 M_{\odot}$, high-mass exponent $\alpha = 2.3$, low-mass exponent $\beta = 1.4$) to match the high mass population from \citet{dezeeuw99}. \\
\indent To estimate the completeness of the high-mass stellar sample in stellar mass we plotted the 93 OB members of Vela OB2 against isochrones from \citet{ekstrom12} in a B-V vs Mv colour-magnitude diagram, using Gaia DR2 parallaxes, which are available for 86 objects (for the remainder we use the median parallax of the sample), see Figure 1. Seven stars with significantly higher B-V values (likely late-type contaminants) were excluded from this sample, as well as 4 stars with Gaia DR2 parallaxes significantly higher or lower than the rest of the sample. Based on the remaining association members relative to the isochrones we estimate the sample completeness limit to be at $\sim2.5 M_{\odot}$ (the turn-over point in the observed luminosity function), with 72 of the 82 remaining OB stars being at least this massive. \\
\indent Ten thousand stellar populations were randomly generated using the \citet{maschberger13} IMF, sampling until 72 stars with $M > 2.5 M_{\odot}$ were produced, and then noting the total number of stars and total mass of the population for each iteration. The median number of stars with $M > 0.1 M_{\odot}$ in a randomly generated population was $1965 \pm 228$ and the median total mass was $1285 \pm 110 M_{\odot}$. \\
\indent \citet{jeffries14} and \citet{prisinzano16} found only 278 PMS likely cluster members in the $0.9$ degree$^2$ area around $\gamma^2$ Vel with a $\sim90\%$ complete sample. Thus, assuming that the whole population of Vela OB2 follows a standard IMF \citep{maschberger13}, this suggests there are $\sim 1700$ association members yet to be identified in the wider area of the association.
\subsection{Searching for the low-mass population of Vela OB2} We compiled a list of all objects within the area around the known OB members of Vela OB2  ($254^{\circ} < l < 271^{\circ}$, $-16^{\circ} < b < 1^{\circ}$) included in both Gaia DR1 and 2MASS catalogues, with a 2MASS photometic quality flag grade of 'AAA', contamination flag grade of '0', and with uncertainties in J, H and K-band magnitudes $< 0.05$ mag . These were then matched to Gaia sources with a matching radius of 0.5 arcseconds, giving a total of 1,028,483 objects.\\
\indent These objects were filtered through selection boxes in G-K vs G and H-K vs J-H diagrams. Fig. 2a shows objects in our sample within the $0.9$ degree$^2$ area around $\gamma^2$ Vel in the G-K vs G diagram (green), with an illustrative 10 Myr isochrone from \citet{baraffe15}, at a distance modulus of 7.76 and reddened by $A_V = 0.131$ mag \citep{jeffries14}. Objects spectroscopically characterised as young stars by \citet{jeffries14} and \citet{prisinzano16} are identified, as are objects spectroscopically confirmed to be contaminating field stars. The previously identified objects form a clear PMS in Fig.2a. The selection box is designed to select this sequence but exclude the extensive contamination at brighter magnitudes. 138 members (red in Fig. 2) and 22 non-members (blue in Fig. 2) from \citet{jeffries14} and \citet{prisinzano16} appear in this selection box.\\
\indent A second selection in the H-K vs J-H diagram was used to filter out any remaining contaminating giants. The H-K vs J-H diagram in Fig. 2b shows the same objects as in the G-K vs G diagram. Distant giants will be reddened away from the PMS stars in this diagram and so we employ the selection box indicated to select just the PMS stars. Combining these two selection methods allows the selection of young, low-mass stars with a much lower rate of contamination. 81 spectroscpically probable members and 4 confirmed non-members appear within both selection boxes. This photometric selection method may exclude PMS stars with substantial accretion disks that may be anomalously red in H-K, however only a very small percentage ($<$10\%) of PMS stars around $\gamma^2$ Velorum were found to have disks and the H-K and J-H colours of most of those would be unaffected \citep{hernandez08}.\\
\indent To extend this selection process to the wider area of Vela OB2, where contaminant levels may differ from that in the immediate area of $\gamma^2$ Velorum, we use data generated by the Besancon Galaxy Model \citep{robin03} to estimate the number of contaminants in our selection boxes (we do this rather than use control fields because it is not clear where one would place a control field given the extended young population in this area). 80 one square degree area samples of model data were taken in the ranges l = 255, 270 and b = -15, 0 deg and a V-I to G-V conversion from \citet{jordi10} was used to produce Gaia-band (G) photometry. Both selection tests were applied to each sample and the numbers of objects passing both tests for each sample give a mean contaminant level of $\sim9.9 \pm 2.6$ deg$^{-2}$, reasonably consistent with the number of spectroscopic non-members found within the $0.9$ degree$^2$ area around $\gamma^2$ Velorum by \citet{jeffries14} and \citet{prisinzano16}. $\sim98\%$ of selected Besancon model objects are at distances $< 200$ pc and are main sequence M-dwarfs in the foreground of Vela OB2. The density map and the structures resolved in it are discussed in more detail in Section 2.3. We calibrate this contamination level with the known contamination rate from \citet{jeffries14}.\\
\indent In order to remove contamination from the older and more distant NGC 2547 cluster (age $\sim$ 35 Myr, \citealt{jeffries05}, distance modulus 8.10, \citealt{naylor02}) we match known cluster members from \citet{jeffries04} to Gaia DR1 and 2MASS and perform the same photometric selection tests. 30 NGC 2547 members pass both tests and thus contribute to contamination in our sample. Accounting for the incompleteness of this sample using the incompleteness estimates in \citet{jeffries04}, we subtract these sources from the relevant areas of the density map in Fig.3. \citet{jeffries04} estimate that $\sim100$ members of NGC 2547 exist beyond the area of their observations. Based on the fraction of \citet{jeffries04} sources that passed our selection tests, we estimate that only 6\% of these will appear in the density map. Their contribution is therefore not significant.\\
\indent PMS evolutionary tracks from \citet{baraffe15} were plotted in a G-K vs G colour-magnitude diagram converted from log $L$ and log $T_{eff}$ using BC values for young stars from Table 6 of \citet{pecaut13}, a V-I to G-V conversion from \citet{jordi10}, alongside our G-K vs G selection box. The selection box was estimated to select stars in the mass ranges $0.45 M_{\odot} > M > 0.17 M_{\odot}$ for a 10 Myr isochrone and $0.39 M_{\odot} > M > 0.16 M_{\odot}$ for a 20 Myr isochrone. The mean number of stars in these mass ranges produced in our randomly generated populations was 815 and 788 stars respectively. \\
\indent We then calculate the numbers of PMS stars shown in the density map. 4882 Gaia+2MASS sources are selected in total, giving 995 PMS stars after applying the background subtraction. The differences between our IMF predictions and density map estimations are reasonable given the uncertainty over the low-mass form of the IMF. IMF uncertainty in the mass range of selected stars and the possibility that some contamination from an older population remains.\\
\subsection{The spatial distributions of high- and low-mass populations}
\begin{figure*}
\begin{center}
\includegraphics[width=380pt]{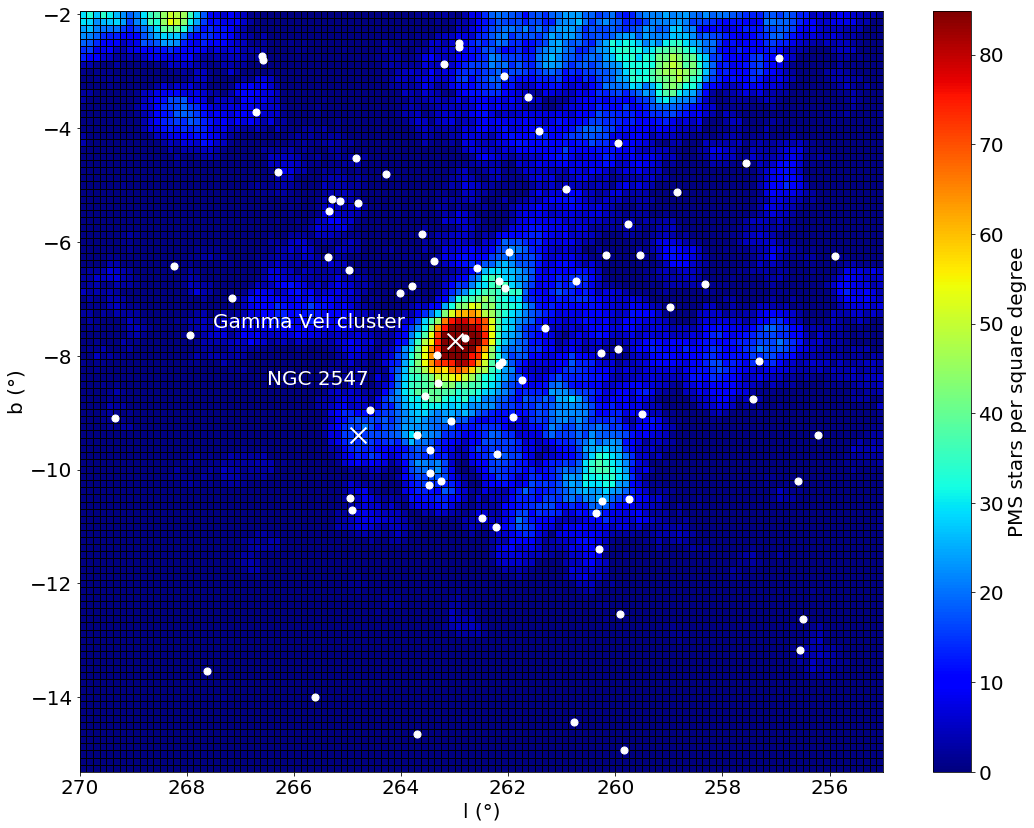}
\setlength{\belowcaptionskip}{-10pt}
\setlength{\textfloatsep}{0pt}
\caption{PMS density map of the area around Vela OB2, with the Besancon model contaminant level subtracted and with the 82 OB members \citep{dezeeuw99} in white. The $\gamma^2$ Vel cluster is clearly visible as the high density area in the centre of the map. We also mark the position of the NGC 2547 cluster which we have subtracted from the density map.}
\label{DensityMap}
\end{center}
\end{figure*}
In our low-mass PMS star density map shown in Fig. 3, the $\gamma^2$ Vel cluster is detected and appears as the region of greatest density of young low-mass stars in the association. After specifically removing the NGC 2547 cluster, there remains a lower-density extended population around $\gamma^2$ Vel and towards the galactic plane. Two other regions of high density appear at (l = 260.25$^\circ$, b = -10.25$^\circ$) and (l = 259.0$^\circ$, b = -3.25$^\circ$), but otherwise the low-mass population appears to be spread sparsely over the Vela OB2 association, albeit with considerable substructure. Notably, the open cluster IC 2395 located at (l = 266.63$^\circ$, b = -3.58$^\circ$) at a distance of $800$ pc and of comparable age to Vela OB2 \citep[$9 \pm 3$ Myr,][]{balog16}, is hardly visible in Figure 3, indicating that our selection method is effective in excluding young objects at much greater distances. \\
\indent It is clear from Figure 3 that the high-mass stars are not preferentially located in areas with high densities of low-mass stars. The two dense regions at (l = 260.25$^\circ$, b = -10.25$^\circ$) and (l = 259.0$^\circ$, b = -3.25$^\circ$) have no high-mass stars within them, whilst the high-mass stars at the lowest latitudes appear isolated from the low-mass population. The latter may be associated with the star forming region RCW38 in the galactic plane, though younger regions near the galactic plane are unlikely contaminants because they will be reddened out of the H-K v J-H selection box, but the former region appears to be a previously-unknown cluster of low-mass stars.\\
\indent In order to estimate a statistical significance to the difference in the distributions of the high- and low-mass populations in Fig.3, a two sample Kolmogorov-Smirnov test was performed on the projected YSO density distributions of the high- and low-mass populations. We calculate the cumulative distribution of YSO densities upon which the high- and low-mass stars are projected. The cumulative distributions calculated are plotted in Figure 4, showing that the high-mass stars are preferentially projected against regions of low density compared to the low-mass stars, with a KS-test P-value of 0.0077. More than$\sim26\%$ of the high-mass stars are projected against cells with low-mass PMS stellar densities of 0.

Finally we note that the methods employed here could lead to the inclusion of stars with notably different asges and distances that still fall in our selection boxes, however this requires these parameters to fall in a relatively narrow range (as evidenced by the absence of IC 2395 in our density map). If a population of stars exists in this parameter space range it could bias our results, but we consider this to be unlikely.

\begin{figure}
\includegraphics[width=\columnwidth]{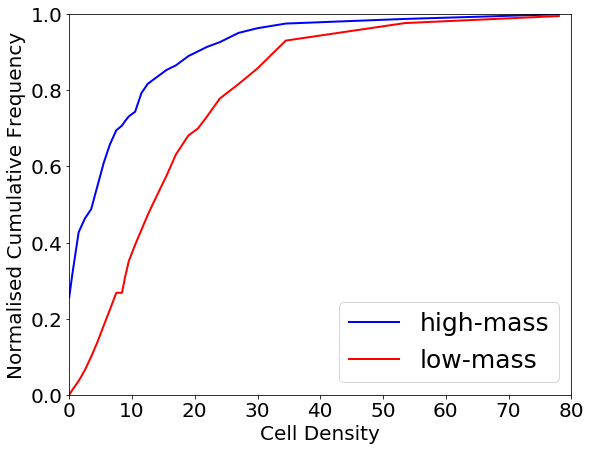}
\setlength{\belowcaptionskip}{-10pt}
\setlength{\textfloatsep}{0pt}
\caption{Cumulative distributions of cell densities of high-mass (blue) and low-mass (red) populations.}
\label{KSbes}

\end{figure}
\section{Discussion} \citet{sacco15} found 15 stars with similar kinematics and age to the $\gamma^2$ Vel cluster members identified by \citet{jeffries14}, but located 2 degrees ($\sim10$ pc) from $\gamma^2$ Velorum, well outside the area studied by \citet{jeffries14}, and speculated that they might belong to a widespread PMS star population associated with Vela OB2. They also commented on the discrepancy between the total mass of the cluster ($\sim100 M_{\odot}$) and the total cluster mass predicted by the mass of $\gamma^2$ Velorum alone \citep[$\sim1000 M_{\odot}$][]{weidner10}. \citet{damiani17} also identified multiple proper motion populations within the higher-mass members of Vela OB2 that they associated with $\gamma^2$ Vel and NGC 2547. \\
\indent Figure 3 shows that our photometric selection has detected a widespread population across the whole area of the association, confirming the speculation of \citet{sacco15}. The agreement between the numbers of PMS stars found in this mass range and the numbers predicted by a standard IMF suggest we have identified the spatial distribution of the low-mass content of Vela OB2. It shows notable concentrations around $\gamma^2$ Vel and the NGC 2547 cluster, but Fig. 4 shows that in general the high mass stars are not closely correlated with the location of the PMS stars.\\
\indent There are two basic possibilities: (a) either the high-mass stars were born in relative isolation - in regions with IMFs that are quite different to the canonical IMF; or (b) the high mass stars have moved away from the low-mass population (in many cases) on timescales of $\sim10$ Myr.\\
\indent The evidence for the success of this photometric selection method in identifying stars in a narrow range of distance and age demonstrates its potential usefulness for identifying stellar populations in other regions, given a pre-existing sample to calibrate the selection criteria. In particular, other nearby associations with samples of known members may be studied in similar detail to this study, allowing us to learn about the distribution and structure of low-mass populations in multiple associations. \\
\indent The second Gaia Data Release provides parallaxes and proper motions for 1.3 billion stars that will allow the spatial and kinematic structure of Vela OB2 to be probed in more detail \citep[e.g.,][]{beccari18,franciosini18}, which will allow a reduction of the two aforementioned scenarios.

\section{Acknowledgments}
We thank the anonymous referee for a careful reading of this paper and helpful suggestions that have improved this work. N.J.W acknowledges an STFC Ernest Rutherford Fellowship (grant number ST/M005569/1). This work has made use of data from the ESA space mission Gaia, processed by the Gaia Data Processing and Analysis Consortium (DPAC). Funding for DPAC has been provided by national institutions, in particular the institutions participating in the Gaia Multilateral Agreement. This research made use of the Simbad and Vizier catalogue access tools (provided by CDS, Strasbourg, France), Astropy \citep{astr13} and TOPCAT \citep{tayl05}. 

\bibliographystyle{mn2e}
\bibliography{LowMassVelaOB2}

\begin{thebibliography}{38}
\expandafter\ifx\csname natexlab\endcsname\relax\def\natexlab#1{#1}\fi

\bibitem[{{Anderson} \& {Francis}(2012)}]{anderson12}
{Anderson} E., {Francis} C., 2012, Astronomy Letters, 38, 331

\bibitem[{{Astropy Collaboration} {et~al}\mbox{.}(2013){Astropy Collaboration},
  {Robitaille}, {Tollerud}, {Greenfield}, {Droettboom}, {Bray}, {Aldcroft},
  {Davis}, {Ginsburg}, {Price-Whelan}, {Kerzendorf}, {Conley}, {Crighton},
  {Barbary}, {Muna}, {Ferguson}, {Grollier}, {Parikh}, {Nair}, {Unther},
  {Deil}, {Woillez}, {Conseil}, {Kramer}, {Turner}, {Singer}, {Fox}, {Weaver},
  {Zabalza}, {Edwards}, {Azalee Bostroem}, {Burke}, {Casey}, {Crawford},
  {Dencheva}, {Ely}, {Jenness}, {Labrie}, {Lim}, {Pierfederici}, {Pontzen},
  {Ptak}, {Refsdal}, {Servillat}, \& {Streicher}}]{astr13}
{Astropy Collaboration} {et~al.}, 2013, \aap, 558, A33

\bibitem[{{Balog} {et~al}\mbox{.}(2016){Balog}, {Siegler}, {Rieke}, {Kiss},
  {Muzerolle}, {Gutermuth}, {Bell}, {Vink{\'o}}, {Su}, {Young}, \&
  {G{\'a}sp{\'a}r}}]{balog16}
{Balog} Z. {et~al.}, 2016, \apj, 832, 87

\bibitem[{{Baraffe} {et~al}\mbox{.}(2015){Baraffe}, {Homeier}, {Allard}, \&
  {Chabrier}}]{baraffe15}
{Baraffe} I., {Homeier} D., {Allard} F., {Chabrier} G., 2015, \aap, 577, A42

\bibitem[{{Beccari} {et~al}\mbox{.}(2018){Beccari}, {Boffin}, {Jerabkova},
  {Wright}, {Kalari}, {Carraro}, {De Marchi}, \& {de Wit}}]{beccari18}
{Beccari} G., {Boffin} H.~M.~J., {Jerabkova} T., {Wright} N.~J., {Kalari}
  V.~M., {Carraro} G., {De Marchi} G., {de Wit} W.-J., 2018, ArXiv e-prints

\bibitem[{{Blaauw}(1964)}]{blaauw64}
{Blaauw} A., 1964, \araa, 2, 213

\bibitem[{{Damiani} {et~al}\mbox{.}(2017){Damiani}, {Prisinzano}, {Jeffries},
  {Sacco}, {Randich}, \& {Micela}}]{damiani17}
{Damiani} F., {Prisinzano} L., {Jeffries} R.~D., {Sacco} G.~G., {Randich} S.,
  {Micela} G., 2017, \aap, 602, L1

\bibitem[{{de Zeeuw} {et~al}\mbox{.}(1999){de Zeeuw}, {Hoogerwerf}, {de
  Bruijne}, {Brown}, \& {Blaauw}}]{dezeeuw99}
{de Zeeuw} P.~T., {Hoogerwerf} R., {de Bruijne} J.~H.~J., {Brown} A.~G.~A.,
  {Blaauw} A., 1999, \aj, 117, 354

\bibitem[{{Ekstr{\"o}m} {et~al}\mbox{.}(2012){Ekstr{\"o}m}, {Georgy},
  {Eggenberger}, {Meynet}, {Mowlavi}, {Wyttenbach}, {Granada}, {Decressin},
  {Hirschi}, {Frischknecht}, {Charbonnel}, \& {Maeder}}]{ekstrom12}
{Ekstr{\"o}m} S. {et~al.}, 2012, \aap, 537, A146

\bibitem[{{Elmegreen} \& {Lada}(1977)}]{elmegreen77}
{Elmegreen} B.~G., {Lada} C.~J., 1977, \apj, 214, 725

\bibitem[{{Franciosini} {et~al}\mbox{.}(2018){Franciosini}, {Sacco},
  {Jeffries}, {Damiani}, {Roccatagliata}, {Fedele}, \&
  {Randich}}]{franciosini18}
{Franciosini} E., {Sacco} G.~G., {Jeffries} R.~D., {Damiani} F.,
  {Roccatagliata} V., {Fedele} D., {Randich} S., 2018, ArXiv e-prints

\bibitem[{{Gaia Collaboration} {et~al}\mbox{.}(2016{\natexlab{a}}){Gaia
  Collaboration}, {Brown}, {Vallenari}, {Prusti}, {de Bruijne}, {Mignard},
  {Drimmel}, {Babusiaux}, {Bailer-Jones}, {Bastian}, \& et~al.}]{gaia16b}
{Gaia Collaboration} {et~al.}, 2016{\natexlab{a}}, \aap, 595, A2

\bibitem[{{Gaia Collaboration} {et~al}\mbox{.}(2016{\natexlab{b}}){Gaia
  Collaboration}, {Prusti}, {de Bruijne}, {Brown}, {Vallenari}, {Babusiaux},
  {Bailer-Jones}, {Bastian}, {Biermann}, {Evans}, \& et~al.}]{gaia16a}
{Gaia Collaboration} {et~al.}, 2016{\natexlab{b}}, \aap, 595, A1

\bibitem[{{Hern{\'a}ndez} {et~al}\mbox{.}(2008){Hern{\'a}ndez}, {Hartmann},
  {Calvet}, {Jeffries}, {Gutermuth}, {Muzerolle}, \& {Stauffer}}]{hernandez08}
{Hern{\'a}ndez} J., {Hartmann} L., {Calvet} N., {Jeffries} R.~D., {Gutermuth}
  R., {Muzerolle} J., {Stauffer} J., 2008, \apj, 686, 1195

\bibitem[{{Hills}(1980)}]{hills80}
{Hills} J.~G., 1980, \apj, 235, 986

\bibitem[{{Jeffries} {et~al}\mbox{.}(2014){Jeffries}, {Jackson}, {Cottaar},
  {Koposov}, {Lanzafame}, {Meyer}, {Prisinzano}, {Randich}, {Sacco},
  {Brugaletta}, {Caramazza}, {Damiani}, {Franciosini}, {Frasca}, {Gilmore},
  {Feltzing}, {Micela}, {Alfaro}, {Bensby}, {Pancino}, {Recio-Blanco}, {de
  Laverny}, {Lewis}, {Magrini}, {Morbidelli}, {Costado}, {Jofr{\'e}},
  {Klutsch}, {Lind}, \& {Maiorca}}]{jeffries14}
{Jeffries} R.~D. {et~al.}, 2014, \aap, 563, A94

\bibitem[{{Jeffries} {et~al}\mbox{.}(2004){Jeffries}, {Naylor}, {Devey}, \&
  {Totten}}]{jeffries04}
{Jeffries} R.~D., {Naylor} T., {Devey} C.~R., {Totten} E.~J., 2004, \mnras,
  351, 1401

\bibitem[{{Jeffries} {et~al}\mbox{.}(2009){Jeffries}, {Naylor}, {Walter},
  {Pozzo}, \& {Devey}}]{jeffries09}
{Jeffries} R.~D., {Naylor} T., {Walter} F.~M., {Pozzo} M.~P., {Devey} C.~R.,
  2009, \mnras, 393, 538

\bibitem[{{Jeffries} \& {Oliveira}(2005)}]{jeffries05}
{Jeffries} R.~D., {Oliveira} J.~M., 2005, \mnras, 358, 13

\bibitem[{{Jordi} {et~al}\mbox{.}(2010){Jordi}, {Gebran}, {Carrasco}, {de
  Bruijne}, {Voss}, {Fabricius}, {Knude}, {Vallenari}, {Kohley}, \&
  {Mora}}]{jordi10}
{Jordi} C. {et~al.}, 2010, \aap, 523, A48

\bibitem[{{Kruijssen}(2012)}]{kruij12}
{Kruijssen} J.~M.~D., 2012, \mnras, 426, 3008

\bibitem[{{Lada} \& {Lada}(2003)}]{lada03}
{Lada} C.~J., {Lada} E.~A., 2003, \araa, 41, 57

\bibitem[{{Lada} {et~al}\mbox{.}(1984){Lada}, {Margulis}, \&
  {Dearborn}}]{lada84}
{Lada} C.~J., {Margulis} M., {Dearborn} D., 1984, \apj, 285, 141

\bibitem[{{Luri} {et~al}\mbox{.}(2018){Luri}, {Brown}, {Sarro}, {Arenou},
  {Bailer-Jones}, {Castro-Ginard}, {de Bruijne}, {Prusti}, {Babusiaux}, \&
  {Delgado}}]{luri18}
{Luri} X. {et~al.}, 2018, ArXiv e-prints

\bibitem[{M.~Cutri {et~al}\mbox{.}(2003)M.~Cutri, F.~Skrutskie, van Dyk,
  A.~Beichman, M.~Carpenter, Chester, Cambresy, Evans, Fowler, Gizis, Howard,
  \& Huchra}]{cutri03}
M.~Cutri R. {et~al.}, 2003

\bibitem[{{Maschberger}(2013)}]{maschberger13}
{Maschberger} T., 2013, \mnras, 429, 1725

\bibitem[{{Naylor} {et~al}\mbox{.}(2002){Naylor}, {Totten}, {Jeffries},
  {Pozzo}, {Devey}, \& {Thompson}}]{naylor02}
{Naylor} T., {Totten} E.~J., {Jeffries} R.~D., {Pozzo} M., {Devey} C.~R.,
  {Thompson} S.~A., 2002, \mnras, 335, 291

\bibitem[{{North} {et~al}\mbox{.}(2007){North}, {Tuthill}, {Tango}, \&
  {Davis}}]{north07}
{North} J.~R., {Tuthill} P.~G., {Tango} W.~J., {Davis} J., 2007, \mnras, 377,
  415

\bibitem[{{Pecaut} \& {Mamajek}(2013)}]{pecaut13}
{Pecaut} M.~J., {Mamajek} E.~E., 2013, \apjs, 208, 9

\bibitem[{{Perryman} {et~al}\mbox{.}(1997){Perryman}, {Lindegren},
  {Kovalevsky}, {Hoeg}, {Bastian}, {Bernacca}, {Cr{\'e}z{\'e}}, {Donati},
  {Grenon}, {Grewing}, {van Leeuwen}, {van der Marel}, {Mignard}, {Murray}, {Le
  Poole}, {Schrijver}, {Turon}, {Arenou}, {Froeschl{\'e}}, \&
  {Petersen}}]{perryman97}
{Perryman} M.~A.~C. {et~al.}, 1997, \aap, 323, L49

\bibitem[{{Pozzo} {et~al}\mbox{.}(2000){Pozzo}, {Jeffries}, {Naylor}, {Totten},
  {Harmer}, \& {Kenyon}}]{pozzo00}
{Pozzo} M., {Jeffries} R.~D., {Naylor} T., {Totten} E.~J., {Harmer} S.,
  {Kenyon} M., 2000, \mnras, 313, L23

\bibitem[{{Prisinzano} {et~al}\mbox{.}(2016){Prisinzano}, {Damiani}, {Micela},
  {Jeffries}, {Franciosini}, {Sacco}, {Frasca}, {Klutsch}, {Lanzafame},
  {Alfaro}, {Biazzo}, {Bonito}, {Bragaglia}, {Caramazza}, {Vallenari},
  {Carraro}, {Costado}, {Flaccomio}, {Jofr{\'e}}, {Lardo}, {Monaco},
  {Morbidelli}, {Mowlavi}, {Pancino}, {Randich}, \& {Zaggia}}]{prisinzano16}
{Prisinzano} L. {et~al.}, 2016, \aap, 589, A70

\bibitem[{{Robin} {et~al}\mbox{.}(2003){Robin}, {Reyl{\'e}}, {Derri{\`e}re}, \&
  {Picaud}}]{robin03}
{Robin} A.~C., {Reyl{\'e}} C., {Derri{\`e}re} S., {Picaud} S., 2003, \aap, 409,
  523

\bibitem[{{Sacco} {et~al}\mbox{.}(2015){Sacco}, {Jeffries}, {Randich},
  {Franciosini}, {Jackson}, {Cottaar}, {Spina}, {Palla}, {Mapelli}, {Alfaro},
  {Bonito}, {Damiani}, {Frasca}, {Klutsch}, {Lanzafame}, {Bayo}, {Barrado},
  {Jim{\'e}nez-Esteban}, {Gilmore}, {Micela}, {Vallenari}, {Allende Prieto},
  {Flaccomio}, {Carraro}, {Costado}, {Jofr{\'e}}, {Lardo}, {Magrini},
  {Morbidelli}, {Prisinzano}, \& {Sbordone}}]{sacco15}
{Sacco} G.~G. {et~al.}, 2015, \aap, 574, L7

\bibitem[{{Taylor}(2005)}]{tayl05}
{Taylor} M.~B., 2005, in Astronomical Society of the Pacific Conference Series,
  Vol. 347, Astronomical Data Analysis Software and Systems XIV, {Shopbell} P.,
  {Britton} M., {Ebert} R., eds., p.~29

\bibitem[{{Weidner} {et~al}\mbox{.}(2010){Weidner}, {Kroupa}, \&
  {Bonnell}}]{weidner10}
{Weidner} C., {Kroupa} P., {Bonnell} I.~A.~D., 2010, \mnras, 401, 275

\bibitem[{{Wright} {et~al}\mbox{.}(2016){Wright}, {Bouy}, {Drew}, {Sarro},
  {Bertin}, {Cuillandre}, \& {Barrado}}]{wright16}
{Wright} N.~J., {Bouy} H., {Drew} J.~E., {Sarro} L.~M., {Bertin} E.,
  {Cuillandre} J.-C., {Barrado} D., 2016, \mnras, 460, 2593

\bibitem[{{Wright} \& {Mamajek}(2018)}]{wright18}
{Wright} N.~J., {Mamajek} E.~E., 2018, \mnras, 476, 381

\end{thebibliography}


\begin{thebibliography}{2}
\expandafter\ifx\csname natexlab\endcsname\relax\def\natexlab#1{#1}\fi

\bibitem[{{Astropy Collaboration} {et~al}\mbox{.}(2013){Astropy Collaboration},
  {Robitaille}, {Tollerud}, {Greenfield}, {Droettboom}, {Bray}, {Aldcroft},
  {Davis}, {Ginsburg}, {Price-Whelan}, {Kerzendorf}, {Conley}, {Crighton},
  {Barbary}, {Muna}, {Ferguson}, {Grollier}, {Parikh}, {Nair}, {Unther},
  {Deil}, {Woillez}, {Conseil}, {Kramer}, {Turner}, {Singer}, {Fox}, {Weaver},
  {Zabalza}, {Edwards}, {Azalee Bostroem}, {Burke}, {Casey}, {Crawford},
  {Dencheva}, {Ely}, {Jenness}, {Labrie}, {Lim}, {Pierfederici}, {Pontzen},
  {Ptak}, {Refsdal}, {Servillat}, \& {Streicher}}]{astr13}
{Astropy Collaboration} {et~al.}, 2013, \aap, 558, A33

\bibitem[{{Taylor}(2005)}]{tayl05}
{Taylor} M.~B., 2005, in Astronomical Society of the Pacific Conference Series,
  Vol. 347, Astronomical Data Analysis Software and Systems XIV, {Shopbell} P.,
  {Britton} M., {Ebert} R., eds., p.~29

\end{thebibliography}
\bsp

\end{document}